\documentclass[conference]{IEEEtran}

\IEEEoverridecommandlockouts

\usepackage{amsmath,amssymb,bm,booktabs,graphicx}
\usepackage[caption=false]{subfig}
\usepackage{algorithm}
\usepackage{algpseudocode}
\usepackage{cite}
\usepackage{url}
\usepackage{balance}
\usepackage{placeins}
\usepackage{hyperref}
\newcommand{\ii}{\mathrm{i}}
\newcommand{\Oh}{\mathcal{O}}
\newcommand{\Rhat}{\widehat{\bm R}}
\newcommand{\Ps}{\bm P_{\rm s}}
\newcommand{\Pn}{\bm P_{\rm n}}
\newcommand{\ket}[1]{\left|#1\right\rangle}
\newcommand{\bra}[1]{\left\langle#1\right|}
\newcommand{\braket}[2]{\left\langle#1\middle|#2\right\rangle}
\newcommand{\norm}[1]{\left\lVert#1\right\rVert}
\newcommand{\CN}{\mathcal{CN}}

\title{Q-SPARSE: Quantum Subspace Projection for Near-Field Angle-Range Spectrum Estimation}

\author{
  Mostafizur~Rahaman~Laskar\textsuperscript{1,*}\thanks{This work was supported by the European Union under the MULTIPLY-6G
project (No. 101293106) and the QUEST-6G project (No. 101292676). G.Fodor was also supported by the Swedish Strategic Research (SSF) grant for the FUS21-0004 SAICOM project.~M.R.Laskar is on sabbatical leave from IBM Research Lab during this work.}
  Sajad~Daei\textsuperscript{1},
  G\'{a}bor~Fodor\textsuperscript{1,2},
  and~Mikael~Skoglund\textsuperscript{1}%
  \\[3pt]
  {\textsuperscript{1}Information Science and Engineering, KTH Royal Institute of Technology, Stockholm, Sweden}\\
  {\textsuperscript{2}Ericsson, Sweden}
}

\begin{document}
\maketitle

\begin{abstract}
Near-field wavefront curvature enables joint angle–range localization, but exploiting it entails repeated covariance-eigenspace tests over a two-dimensional spherical manifold. 
We propose a quantum subspace-projection algorithm (Q-SPARSE) for super-resolution estimation of parameters in the near field, employing spectral signature similar to MUSIC. 
Q-SPARSE prepares each spherical steering hypothesis as a quantum state, applies quantum phase estimation (QPE) to covariance-generated evolution, and converts the resolved spectral mass into a localization spectrum. The proposed method does not need the full classical preparation of the eigenstates, which is a major bottleneck of the QPE algorithm, the main subroutine in quantum variants of MUSIC algorithms. Under an ideal signal–noise partition, its score equals the classical signal-subspace overlap exactly and thus preserves the maximizer of normalized near-field MUSIC. For a finite $q$ number of qubit phase registers, we derive the score from the complete QPE kernel, and propose a nearest-bin selection scheme with soft-thresholding based on  Marchenko–Pastur-calibrated logistic weighting decision criteria. Numerical results on IBM's QISKIT software platform are demonstrated for near-field array processing.
\end{abstract}


\section{Introduction}

Near-field wavefront curvature turns an electrically large array into a joint angular and ranging aperture. At millimeter-wave and sub-terahertz frequencies, practical targets may lie inside the radiative near field, where the array response is spherical and explicitly range dependent. Reliable near-field localization preserve the spherical manifold and estimate angle and range jointly \cite{Wang2023NFISAC,Cong2024NFISAC,Cui2022Polar,Daei2025ISAC,Daei2025Boundary}. Subspace methods provide super-resolution estimates of these parameters at a high computational cost. One of the best known classical subspace algorithms, called multiple signal classification (MUSIC) localizes sources by testing candidate steering vectors against the estimated covariance noise subspace \cite{Schmidt1986MUSIC}, while near-field MUSIC employs spherical steering vectors and searches a two-dimensional angle-range domain \cite{Zhang2018NearFieldMUSIC}. Lifted convex and continuous off-grid formulations further address the nonlinear geometry and discretization effects of the near-field manifold \cite{daei2026convexity,daei2026living}. Despite their different formulations, these methods share a demanding computational structure: the sample covariance needs to be spectrally processed, after which a potentially large number of angle-range hypotheses has to be evaluated. This cost grows with both the array dimension, target size, and the resolution of the polar search grid. To overcome the computational challenges, quantum assisted MUSIC algorithm (often called Q-MUSIC) have been proposed in the literature, where for sparse-systems computational complexity may get improved \cite{meng2020quantum,laskar2022eigentcom}.


Quantum phase estimation (QPE) estimates the eigen-phases of a unitary with its eigenstates \cite{Cleve1998QPE}, given the prepared eigenstates (or eigenvectors) of a unitary matrix\cite{Lloyd2014QPCA}. Existing quantum MUSIC approaches have primarily considered far-field direction finding and covariance-eigenspace labeling \cite{meng2020quantum}; related quantum-assisted radar-communication studies have addressed detection and estimation tasks \cite{Laskar2023QRADCOM,Naskar2026QRadCom}. A direct extension to the near field is nontrivial for two reasons. First, every hypothesis is a spherical quantum state indexed jointly by angle and range. Second, finite QPE resolution redistributes each eigen-component across multiple phase bins, so the complete leakage kernel can change the relative heights of closely competing peaks in the two-dimensional localization spectrum. In addition, classical preparation of eigenstates in Q-MUSIC \cite{meng2020quantum} is costly and needs to be addressed.  

Given this background, we propose \emph{Quantum Subspace Projection for Angle-Range Spectrum Estimation} (Q-SPARSE) algorithm. Based on polar-grid hypothesis, Q-SPARSE prepares the corresponding spherical steering state, inspired by polar MUSIC, but prepared in a quantum basis set.  The main contributions are given as follows:
\begin{itemize}
\item \textbf{Spectral-query formulation and exact equivalence:}
We propose near-field angle-range localization as a quantum spectral-membership problem and show that, under an ideal signal-noise partition, the Q-SPARSE score is similar to the normalized MUSIC signal-subspace score. The computational cost in this method can be less than that of the standard MUSIC approaches, with some assumptions. 

\item \textbf{Finite-resolution characterization:}
We propose criteria for soft-thresholding in the proposed Q-SPARSE algorithm for range and angle estimation in the near-field array processing by quantum subspace method. It avoids full knowledge of the classical preparation of the eigenstates, which is a bottleneck in Q-MUSIC\cite{meng2020quantum}. 

\item \textbf{Resource-aware numerical validation:}
We have demonstrated numerical results in the QISKIT script using the IBM quantum simulator for illustrating that the method works on real system, and especially the application of the bin-selection mechanism proposed in the algorithm. The results are tested and compared with baselines. 
\end{itemize}


\section{Near-Field Model and Subspace Criterion}
\subsection{Spherical Array and Snapshot Model}
We consider a centered $N$-sensor uniform linear array with spacing $d=\lambda/2$ and positions $x_n=(n-(N-1)/2)d$, $n=0,\ldots,N-1$. A source at range $r$ and broadside angle $\theta$ lies at distance
\begin{align}
d_n(r,\theta)&=\sqrt{r^2+x_n^2-2rx_n\sin\theta}, \\
[\bm a(r,\theta)]_n&=\frac{1}{\sqrt N}
\exp\!\left[-\ii\frac{2\pi}{\lambda}(d_n(r,\theta)-r)\right].
\label{eq:steering_conf}
\end{align}
where $\lambda$ is the wavelength. We remove the common phase at the array center, and the factor $N^{-1/2}$ gives $\norm{\bm a}_2=1$. We write physical angle as $\varphi=90^\circ-\theta$. The Fresnel expansion
\begin{align}
d_n-r=-x_n\sin\theta+\frac{x_n^2\cos^2\theta}{2r}
+\Oh\!\left(\frac{|x_n|^3}{r^2}\right)
\end{align}
separates the linear direction term from the quadratic range term. When the quadratic term is negligible, the manifold approaches its far-field form and range cannot be inferred from a narrowband spatial snapshot alone.

For $K$ uncorrelated narrowband sources and $L$ snapshots, we use
\begin{align}
\bm Y&=\bm A\bm S+\bm W, &
\Rhat&=L^{-1}\bm Y\bm Y^H,
\label{eq:model_conf}
\end{align}
where $\bm A=[\bm a(r_1,\theta_1),\ldots,\bm a(r_K,\theta_K)]$, $[\bm W]_{n\ell}\sim\CN(0,\sigma_n^2)$, and $K<\min(N,L)$. The latter condition leaves a nonempty sample noise subspace; increasing the phase-register width cannot replace missing spatial rank or snapshots.

\subsection{Signal-Projector Form of Near-Field MUSIC}
For analysis, write $\Rhat=\sum_{j=1}^{N}\lambda_j\ket{u_j}\bra{u_j}$ with $\lambda_1\geq\cdots\geq\lambda_N\geq0$. If the leading $K$ eigenvectors span the sample signal subspace, define
\begin{align}
\Ps&=\sum_{j=1}^{K}\ket{u_j}\bra{u_j}, &
\Pn&=\bm I-\Ps .
\label{eq:projectors_conf}
\end{align}
Normalized near-field MUSIC evaluates
\begin{align}
P_{\rm MU}(r,\theta)
&=\left[\bra{a(r,\theta)}\Pn\ket{a(r,\theta)}\right]^{-1} \\
&=\left[1-\bra{a(r,\theta)}\Ps\ket{a(r,\theta)}\right]^{-1}.
\label{eq:music_identity_conf}
\end{align}
Hence the MUSIC maxima coincide with the maxima of
\begin{align}
S_{\rm sig}(r,\theta)
&=\bra{a(r,\theta)}\Ps\ket{a(r,\theta)}
=\sum_{j=1}^{K}|\braket{u_j}{a(r,\theta)}|^2 .
\label{eq:signal_score_conf}
\end{align}
This identity provides the target quantity for Q-SPARSE. It also fixes the interpretation: an ideal quantum score cannot be statistically better than MUSIC when both algorithms use the same covariance, manifold, and eigenspace partition.

\section{Q-SPARSE Algorithm}
\subsection{Covariance Spectrum and Soft Decision}
Choose $\alpha>\norm{\Rhat}_2$ and define
\begin{align}
\bm U_R&=\exp(\ii2\pi\Rhat/\alpha), &
\phi_j&=\lambda_j/\alpha\in[0,1).
\label{eq:unitary_conf}
\end{align}
For a candidate state $\ket{a}=\sum_j c_j\ket{u_j}$, $c_j=\braket{u_j}{a}$, QPE labels the covariance eigencomponents without requiring the circuit to receive the eigenvectors as classical input.

We assign phase-bin mass with the smooth spectral weight
\begin{align}
h_{\rm S}(\lambda)&=\left[1+\exp\!\left(-\frac{\lambda-\lambda_+}{\delta}\right)\right]^{-1}, \\
\lambda_+&=\sigma_n^2\left(1+\sqrt{N/L}\right)^2,\qquad
\delta=\beta\lambda_+ .
\label{eq:soft_rule_conf}
\end{align}
Here $\lambda_+$ is the asymptotic upper Marchenko-Pastur edge for spatially white noise, and $\beta$ controls the transition width. We use this edge as a reference rather than a finite-sample false-alarm guarantee. Colored noise requires prewhitening, an estimated noise covariance, or another spectral boundary. As $\beta\downarrow0$, $h_{\rm S}$ approaches the hard signal-noise decision away from the edge.

\subsection{Finite-QPE Score and Measurement Statistics}
With $Q=2^q$, QPE assigns eigenphase $\phi_j$ to phase bin $m$ with probability
\begin{align}
\kappa_q(m\mid\phi_j)=
\frac{\sin^2[\pi Q(\phi_j-m/Q)]}
{Q^2\sin^2[\pi(\phi_j-m/Q)]}.
\label{eq:qpe_kernel_conf}
\end{align}
where we use the continuous value at a removable singularity. The phase histogram for a polar hypothesis is
\begin{align}
 p_m(r,\theta)
 &=\sum_{j=1}^{N}|c_j(r,\theta)|^2\kappa_q(m\mid\phi_j).
\label{eq:phase_hist_conf}
\end{align}
We may weight a measured histogram classically or implement a phase-controlled flag rotation with amplitudes $\sqrt{1-h_m}$ and $\sqrt{h_m}$, where $h_m=h_{\rm S}(\alpha m/Q)$. Both routes give the finite-$q$ Q-SPARSE score
\begin{align}
S_{\rm Q}^{(q)}(r,\theta)
&=\sum_{m=0}^{Q-1}h_{\rm S}(\alpha m/Q)p_m(r,\theta)=\sum_{j=1}^{N}|c_j|^2\widetilde h_j^{(q)}, \\
\widetilde h_j^{(q)}
&=\sum_{m=0}^{Q-1}h_{\rm S}(\alpha m/Q)\kappa_q(m\mid\phi_j).
\label{eq:soft_score_conf}
\end{align}
The QPE kernel filters the logistic response rather than merely rounding each eigenvalue. Its eigenvalue-bin spacing is $\Delta\lambda=\alpha/2^q$, but \eqref{eq:soft_score_conf} retains leakage beyond the nearest bin.

If $X\sim\operatorname{Binomial}(M,S_{\rm Q}^{(q)})$, then $\widehat S_{\rm Q}=X/M$ is conditionally unbiased and
\begin{align}
\operatorname{Var}(\widehat S_{\rm Q}\mid\Rhat)
&=\frac{S_{\rm Q}^{(q)}(1-S_{\rm Q}^{(q)})}{M}\leq\frac{1}{4M}, \\
\Pr\{|\widehat S_{\rm Q}-S_{\rm Q}^{(q)}|\geq\epsilon\}
&\leq2\exp(-2M\epsilon^2).
\label{eq:shot_conf}
\end{align}
Finite snapshots perturb $\Rhat$ before QPE, finite $q$ changes the deterministic spectral response, and finite $M$ adds conditional sampling variation. 

\begin{algorithm}[!htb]
\caption{Q-SPARSE range-angle estimation}
\label{alg:ss_conf}
\begin{algorithmic}[1]
\Require $\bm Y$ or coherent covariance access; grid $\mathcal G$; $q$, $M$, $\beta$; requested peak count $K$.
\Ensure Separated range-angle estimates.
\State Form or encode $\Rhat$; select $\alpha>\norm{\Rhat}_2$.
\State Set $\lambda_+$ and $h_{\rm S}$ from \eqref{eq:soft_rule_conf}.
\ForAll{$(r,\theta)\in\mathcal G$}
    \State Prepare $\ket{a(r,\theta)}$ using \eqref{eq:steering_conf}.
    \State Apply $q$-qubit QPE to $\exp(\ii2\pi\Rhat/\alpha)$.
    \State Weight phase bin $m$ by $h_{\rm S}(\alpha m/2^q)$.
    \State Estimate $\widehat S_{\rm Q}^{(q)}(r,\theta)$ from $M$ shots.
\EndFor
\State Apply two-dimensional nonmaximum suppression.
\State Return the $K$ strongest separated peaks or those above a calibrated threshold.
\end{algorithmic}
\end{algorithm}
\FloatBarrier

\subsection{Computational Complexity}
Let $N$, $L$, $K$, and $G$ denote the numbers of sensors,
snapshots, sources, and searched range-angle points, respectively.
For Quantum MUSIC, $\epsilon_{\rm M}$ denotes the approximation
accuracy used in \cite{meng2020quantum}. For Q-SPARSE, let
$\epsilon_{\rm Q}$ denote both the normalized eigenphase resolution
and the required additive score accuracy. This choice gives
$q=\Oh(\log(1/\epsilon_{\rm Q}))$ phase qubits and
$M=\Oh(\epsilon_{\rm Q}^{-2})$ shots per grid point.

\begin{table}[!htb]
\centering
\caption{Dominant estimation complexity for a $G$-point search.}
\label{tab:complexity_conf}
\renewcommand{\arraystretch}{1.07}
\footnotesize
\begin{tabular}{@{}p{0.38\columnwidth}p{0.56\columnwidth}@{}}
\toprule
Method & Estimation complexity \\
\midrule
NF-MUSIC
\cite{Schmidt1986MUSIC,Zhang2018NearFieldMUSIC}
& $\Oh\!\left(N^3+GN(N-K)\right)$ \\

Loaded MVDR \cite{capon1969high}
& $\Oh\!\left(N^3+GN^2\right)$ \\

Quantum MUSIC
\cite{meng2020quantum}
& $\Oh\!\left(
\kappa_S^{3/2}
\operatorname{poly}(K\log(NKL))/
\epsilon_{\rm M}^{6}\right)$ \\

Q-SPARSE (structured)
& $\widetilde{\Oh}\!\left(
G\epsilon_{\rm Q}^{-3}
\operatorname{poly}(\log N)\right)$ \\

Q-SPARSE (dense)
& $\Oh\!\left(
GN^2\epsilon_{\rm Q}^{-2}
\log(1/\epsilon_{\rm Q})\right)$ \\
\bottomrule
\end{tabular}
\end{table}

Here, $N-K$ is the noise-subspace dimension and
$
\kappa_S=
\frac{\lambda_{\max}(\bm S\bm S^H)}
{\lambda_{\min}^{+}(\bm S\bm S^H)}
$
is the effective condition number used by Quantum MUSIC, where
$\lambda_{\min}^{+}$ is the smallest retained positive eigenvalue.
The structured Q-SPARSE bound assumes that steering-state preparation
and normalized covariance block encoding require
$\operatorname{poly}(\log N)$ operations. If the block encoding uses
an $s$-sparse covariance oracle, where $s$ is the maximum number of
nonzero entries in any row or column, the structured cost becomes
$\widetilde{\Oh}(Gs\epsilon_{\rm Q}^{-3}
\operatorname{poly}(\log N))$.

For the hard projector, however, the eigenvalue resolution
need to satisfy
$\alpha\epsilon_{\rm Q}<(\lambda_K-\lambda_{K+1})/2$; the soft
projector Q-SPARSE requires sufficient resolution relative to its
transition width $\delta$. The poly-logarithmic factor depends
on the state encoding, covariance block encoding, phase arithmetic,
and spectral-weight implementation. The dense bound assumes generic
$\Oh(N^2)$ synthesis of each controlled covariance power and is
reported before routing overhead. Consequently, these conditional
operational bounds do not establish an unconditional quantum speedup.

\section{Numerical Results}
\label{sec:results}
We set \(f_c=28\) GHz, \(d=\lambda/2\), \(r\in[0.15,1]\) m, and \(\varphi\in[0^\circ,180^\circ]\), with grid spacings of (0.025) m and \(0.5^\circ\), respectively. We draw target locations uniformly, generate circularly symmetric complex Gaussian source signals, scale the noiseless signals to achieve the desired signal-to-noise ratio (SNR), and add unit-variance additive white Gaussian noise (AWGN) at the sensors.. We use $q=10$, $M=10^4$, $\beta=0.25$, and $\alpha=1.05\lambda_{\max}(\Rhat)$. Each operating point contains 2000 scenes. The sweeps use the exact finite-$q$ kernel; a separate QISKIT-Aer simulator provides the transpiled-circuit results.

After nonmaximum suppression, a Hungarian assignment pairs targets and estimates. We report range and angle Normalized Mean Squared Error (NMSE) separately:
\begin{align}
{\rm NMSE}_{x}
=\frac{\sum_{t,k}(\widehat x_{k,t}-x_{k,t})^2}
{\sum_{t,k}x_{k,t}^2},\qquad x\in\{r,\varphi\}.
\label{eq:nmse_conf}
\end{align}
For detection, both errors must lie within $0.08$ m and $2^\circ$; a pair missing either gate counts as a missed detection, so \eqref{eq:nmse_conf} is computed only over gated pairs.

\begin{figure}[!htb]
\centering
\includegraphics[width=.36\textwidth]{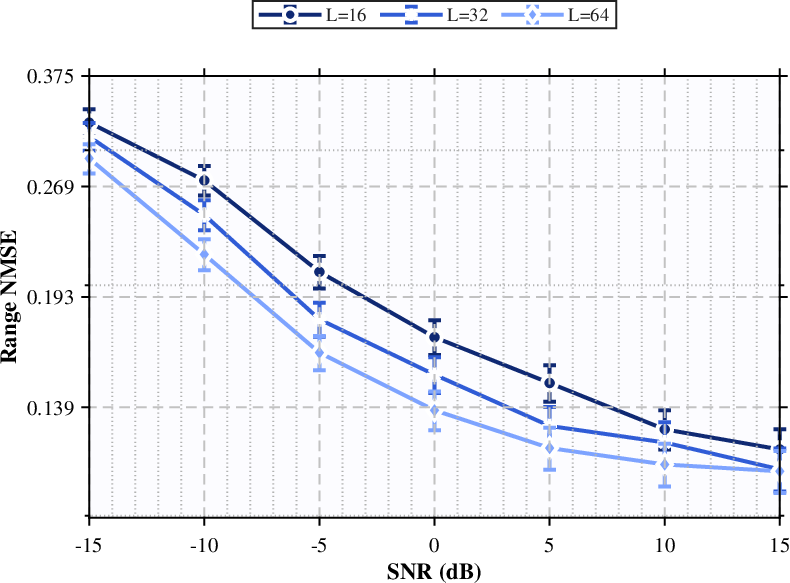}
\caption{Range NMSE versus SNR, $K=2$, for $L\in\{16,32,64\}$ snapshots.}
\label{fig:rnmse}
\end{figure}
\FloatBarrier

\begin{figure}[!htb]
\centering
\includegraphics[width=.36\textwidth]{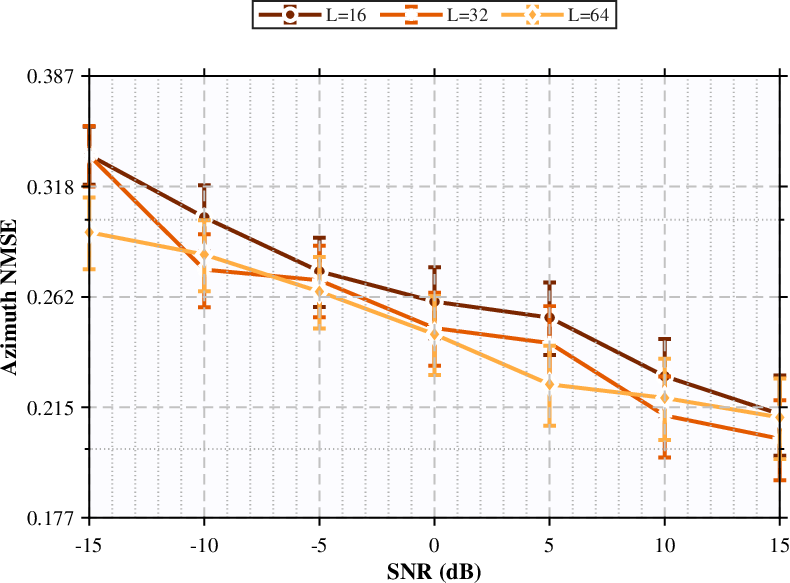}
\caption{Azimuth angle NMSE versus SNR, $K=2$, for $L\in\{16,32,64\}$ snapshots.}
\label{fig:anmse}
\end{figure}
\FloatBarrier

Panels in Figs.~\ref{fig:rnmse}-\ref{fig:anmse} have overlapping intervals across snapshot counts, so no single value of $L$ is uniformly best over the tested SNR range. Two effects overlap here. First, $\lambda_+$ in \eqref{eq:soft_rule_conf} already depends on $L$, so a larger snapshot count sharpens the noise edge and narrows the transition width $\delta=\beta\lambda_+$ at the same time as it improves the sample covariance; part of the expected snapshot gain is absorbed into where the soft threshold sits rather than showing up as a clean NMSE improvement. Second, once the SNR is high enough that both targets clear the calibrated grid resolution, the residual NMSE floor reflects the $0.025$ m and $0.5^\circ$ grid spacing rather than the covariance estimate, which is why the curves flatten at the higher end of the sweep instead of continuing to fall. The occasional crossing between the $L=16$ and $L=32$ curves is within what 2000-scene Monte Carlo sampling can produce given the overlapping confidence intervals, and is not read as a violation of any monotonicity claim; the more informative pattern is that all three snapshot counts converge as $\text{SNR}\rightarrow15$ dB.

\begin{figure}[!htb]
\centering
\includegraphics[width=.36\textwidth]{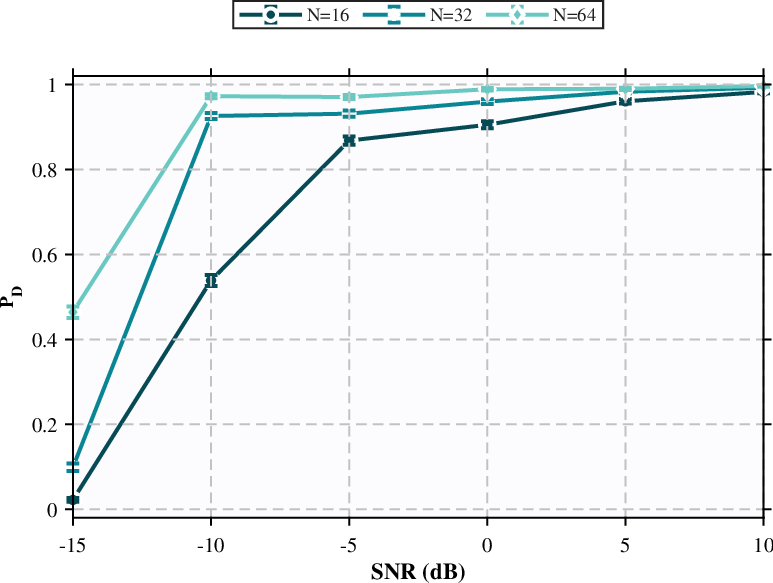}
\caption{Detection probability $P_{\rm D}$ versus $N$, $K=3$, at a noise-only false-alarm calibration of $10^{-2}$, $L=64$.}
\label{fig:pdn}
\end{figure}
\FloatBarrier

\begin{figure}[!htb]
\centering
\includegraphics[width=.36\textwidth]{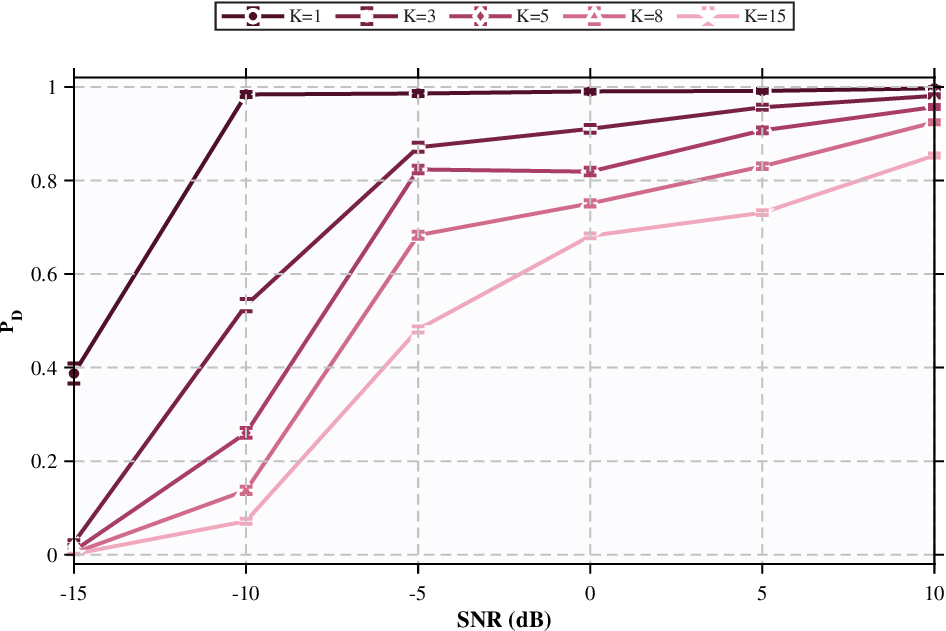}
\caption{Detection probability $P_{\rm D}$ versus $K$, $N=16$, at a noise-only false-alarm calibration of $10^{-2}$, $L=64$.}
\label{fig:pdk}
\end{figure}
\FloatBarrier

Figs.~\ref{fig:pdn}-\ref{fig:pdk} show better low-SNR recovery with a larger array and lower recovery under heavier target loading, consistent with \eqref{eq:signal_score_conf}: a larger $N$ sharpens the steering-vector overlaps, letting the soft score in \eqref{eq:soft_score_conf} clear the false-alarm-calibrated threshold $\gamma_N$ at lower SNR. Since $\gamma_N$ is calibrated separately for each $N$ from the noise-only score distribution, the gain in Fig.~\ref{fig:pdn} is an aperture effect on the signal side rather than an easing of the false-alarm control itself. In Fig.~\ref{fig:pdk}, a larger $K$ at fixed $N=16$ reduces the noise-subspace dimension available to reject spurious peaks, so detection degrades with loading even though the array and calibration are unchanged; this foreshadows the loading-stress reading we give the $K=10$ benchmark below.

\begin{figure}[!htb]
\centering
\includegraphics[width=.36\textwidth]{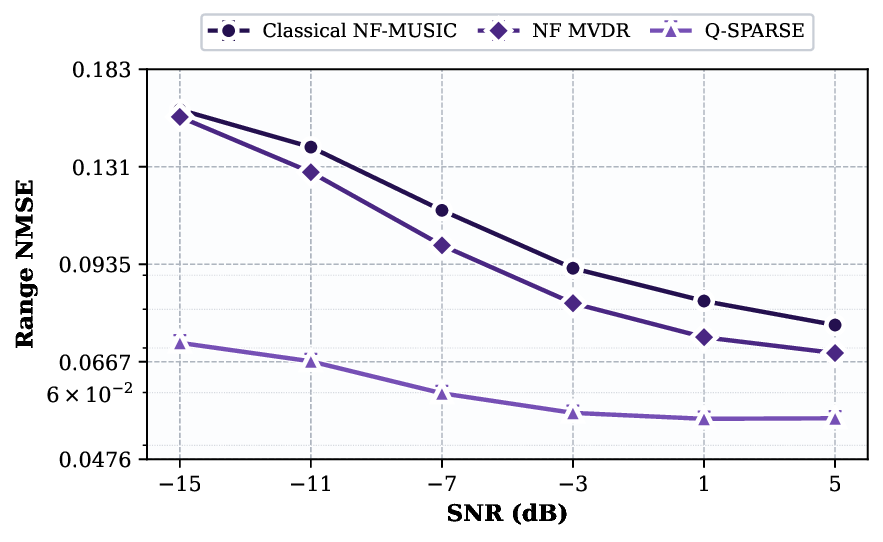}
\caption{Range NMSE for Q-SPARSE, near-field MUSIC, and loaded MVDR on identical scenes, $N=16$, $K=10$, $L=64$.}
\label{fig:benchmark}
\end{figure}
\FloatBarrier

Fig.~\ref{fig:benchmark} compares Q-SPARSE with near-field MUSIC and loaded MVDR on identical scenes. Its range NMSE falls from about $7.6\times10^{-2}$ at $-15$ dB to $5.2\times10^{-2}$ near $0$ dB and, beyond it remains near $(5.3$-$5.7)\times10^{-2}$ approximately. The gap comes from the logistic covariance weighting, which also has a classical matrix-function realization and is not evidence of an accuracy gain over the classical baselines, since \eqref{eq:signal_score_conf} gives no reason to expect one under a hard signal-noise split. The azimuth angle benchmark shows no method with the smallest NMSE throughout the SNR interval, and forced return of ten peaks with $K=10$, $N=16$ makes this a loading stress test rather than a consistency experiment. Loaded MVDR sits closer to Q-SPARSE than classical NF-MUSIC across most of the sweep, plausibly because its diagonal loading damps low-eigenvalue components in a way related to, though not identical to, the logistic weight in \eqref{eq:soft_rule_conf}.

A separate transpiled dense-unitary validation circuit, used only as a small-array check, raises circuit depth from about $4.8\times10^3$ to $1.0\times10^4$ and CX count from $1.2\times10^3$ to $2.5\times10^3$ as $q$ widens from $5$ to $10$, with gate-error probabilities $10^{-4}$ and $10^{-3}$ and readout error $2\times10^{-3}$. Both quantities grow roughly linearly with $q$ since each extra phase qubit adds one more controlled power of the dense covariance unitary before the inverse QFT; at $N=16$ the controlled-unitary synthesis, sets the two-qubit gate count. The computational cost is shown in Table~\ref{tab:complexity_conf}. Figs.~\ref{fig:rnmse}-\ref{fig:benchmark} considered AWGN, finite-QPE resolution, and shot sampling to show robustness of the proposed algorithm under various perturbations.

\section{Conclusion}
We propose Q-SPARSE, a quantum subspace-projection framework for joint near-field angle–range estimation that avoids explicit covariance-eigenvector preparation in the quantum phase estimation routine. Under efficient state-preparation and structured covariance-access assumptions, Q-SPARSE achieves accuracy comparable to that of classical MUSIC at substantially lower computational complexity. We further introduce threshold-based spectral-signature selection using a Marchenko-Pastur-based soft-decision criterion. Numerical results validate the resulting localization spectrum and quantify the effects of finite phase resolution, measurement shot noise, and observation noise over randomized target positions.


\balance
\bibliographystyle{IEEEtran}
\bibliography{references}

@article{Schmidt1986MUSIC,
  author={R. O. Schmidt}, title={Multiple emitter location and signal parameter estimation}, journal={IEEE Transactions on Antennas and Propagation}, volume={34}, number={3}, pages={276--280}, month=mar, year={1986}, doi={10.1109/TAP.1986.1143830}}

@article{Zhang2018NearFieldMUSIC,
  author={X. Zhang and W. Chen and W. Zheng and Z. Xia and Y. Wang}, title={Localization of Near-Field Sources: A Reduced-Dimension MUSIC Algorithm}, journal={IEEE Communications Letters}, volume={22}, number={7}, pages={1422--1425}, month=jul, year={2018}, doi={10.1109/LCOMM.2018.2837049}}

@article{Cui2022Polar,
  author={M. Cui and L. Dai}, title={Channel Estimation for Extremely Large-Scale MIMO: Far-Field or Near-Field?}, journal={IEEE Transactions on Communications}, volume={70}, number={4}, pages={2663--2677}, month=apr, year={2022}, doi={10.1109/TCOMM.2022.3146400}}

@article{Wang2023NFISAC,
  author={Z. Wang and X. Mu and Y. Liu}, title={Near-Field Integrated Sensing and Communications}, journal={IEEE Communications Letters}, volume={27}, number={8}, pages={2048--2052}, month=aug, year={2023}, doi={10.1109/LCOMM.2023.3280132}}

@article{Cong2024NFISAC,
  author={J. Cong and C. You and J. Li and L. Chen and B. Zheng and Y. Liu and W. Wu and Y. Gong and S. Jin and R. Zhang}, title={Near-Field Integrated Sensing and Communication: Opportunities and Challenges}, journal={IEEE Wireless Communications}, volume={31}, number={6}, pages={162--169}, month=dec, year={2024}, doi={10.1109/MWC.002.2400033}}

@inproceedings{Daei2025ISAC,
  author={S. Daei and A. Zamani and S. Chatterjee and M. Skoglund and G. Fodor}, title={Near-Field ISAC in 6G: Addressing Phase Nonlinearity via Lifted Super-Resolution}, booktitle={Proc. IEEE Int. Conf. Acoustics, Speech and Signal Processing (ICASSP)}, year={2025}, doi={10.1109/ICASSP49660.2025.10888339}}

@inproceedings{daei2026convexity,
  author    = {Daei, Sajad and Fodor, G{\'a}bor and Skoglund, Mikael},
  title     = {Convexity Meets Curvature: Lifted Near-Field Super-Resolution},
  booktitle = {Proceedings of the European Signal Processing Conference
               (EUSIPCO)},
  year      = {2026},
  note      = {Also available as arXiv:2602.14063}
}

@inproceedings{daei2026living,
  author    = {{S. Daei, G. Fodor, and M. Skoglund}},
  title     = {Living Off the Grid: Continuous Range-Angle Super-Resolution
               for Near-Field XL-MIMO},
  booktitle = {Proceedings of the IEEE International Workshop on Signal
               Processing Advances in Wireless Communications (SPAWC)},
  year      = {2026},
  note      = {Also available as arXiv:2604.10234}
}

@article{meng2020quantum,
  title={Quantum algorithm for multiple signal classification},
  author={Meng, Fan-Xu and Yu, Xu-Tao and Zhang, Zai-Chen},
  journal={Physical Review A},
  volume={101},
  number={1},
  pages={012334},
  year={2020},
  publisher={APS}
}

@article{capon1969high,
  title={High-resolution frequency-wavenumber spectrum analysis},
  author={Capon, Jack},
  journal={Proceedings of the IEEE},
  volume={57},
  number={8},
  pages={1408--1418},
  year={1969},
  publisher={IEEE}
}

@article{laskar2022eigentcom,
  title={Eigen-spectrum estimation and source detection in a massive sensor array based on quantum assisted Hamiltonian simulation framework},
  author={Laskar, Mostafizur Rahaman and Mondal, Subhadeep and Dutta, Amit Kumar},
  journal={IEEE Transactions on Communications},
  volume={70},
  number={6},
  pages={4013--4025},
  year={2022},
  publisher={IEEE}
}

@inproceedings{Daei2025Boundary,
  author={S. Daei and G. Fodor and M. Skoglund}, title={When Near Becomes Far: From Rayleigh to Optimal Near-Field and Far-Field Boundaries}, booktitle={Proc. IEEE Global Communications Conference (GLOBECOM)}, pages={3586--3592}, year={2025}, doi={10.1109/GLOBECOM59602.2025.11431789}}

@article{Cleve1998QPE,
  author={R. Cleve and A. Ekert and C. Macchiavello and M. Mosca}, title={Quantum Algorithms Revisited}, journal={Proceedings of the Royal Society of London. Series A: Mathematical, Physical and Engineering Sciences}, volume={454}, number={1969}, pages={339--354}, year={1998}, doi={10.1098/rspa.1998.0164}}

@article{Lloyd2014QPCA,
  author={S. Lloyd and M. Mohseni and P. Rebentrost}, title={Quantum Principal Component Analysis}, journal={Nature Physics}, volume={10}, pages={631--633}, year={2014}, doi={10.1038/nphys3029}}

@inproceedings{Laskar2023QRADCOM,
  author={M. R. Laskar and S. Naskar and A. K. Dutta}, title={QRADCOM: Quantum Assisted Framework for Joint Detection and Estimation in Radar Communications}, booktitle={Proc. IEEE 97th Vehicular Technology Conference (VTC2023-Spring)}, pages={1--7}, year={2023}, doi={10.1109/VTC2023-Spring57618.2023.10200135}}

@article{Naskar2026QRadCom,
  title={QRadCom: Quantum-assisted source and symbol detection framework in a radar communication system with low-computational complexity},
  author={Naskar, Soumita and Laskar, Mostafizur Rahaman and Dutta, Amit Kumar},
  journal={IEEE Transactions on Cognitive Communications and Networking},
  volume={12},
  pages={912--927},
  year={2025},
  publisher={IEEE}
}

\end{document}